# Blockchain Smart Contract Threat Detection Technology Based on Symbolic Execution

Chang Chu
chuc23@mails.tsinghua.edu.cn

**Abstract**
The security of smart contracts, which are an important part of blockchain technology, has attracted much attention. In particular, reentrancy vulnerability, which is hidden and complex, poses a great threat to smart contracts. In order to improve the existing detection methods, which exhibit low efficiency and accuracy, in this paper, we propose a smart contract threat detection technology based on symbolic execution. In this method, first, the recursive descent algorithm is used to recover the basic blocks of contract code and control flow diagram, and static type inference is performed for static single assignment (SSA) variables. Then, the control flow diagram is encoded into constrained horn clause (CHC) constraints in combination with the symbolic execution technology. Model checking is conducted for the generated constraints using an automatic theorem prover based on the abstraction refinement technique for fast static detection of common security threats in smart contracts. Compared with existing detection methods, the method proposed in this paper allows the detection of both the checks-effects-interactions pattern and the vulnerability in relation to reentrant locks. It can simulate the state changes of reentrant locks as well as other global variables in multiple recursive transactions. The experimental results show that this method significantly increases both detection efficiency and accuracy, improving the security of smart contracts.

## 1 Introduction

### 1.1 Background and Significance of the Study

The popularization and application of blockchain technology have led to a rapid growth in the deployment of smart contracts. However, the execution process of smart contracts is complex and uncertain because they are vulnerable to various attacks [1] that may result in smart contract tampering and theft, causing significant losses to users and platforms. Therefore, in order to ensure the security and stable operation of the blockchain, improving the security of smart contracts is critical.

### 1.2 History and Current Status of Local and Global Studies

Here, model checking refers to the use of mathematical logic to abstract the execution behavior of programs and assess the constraint model in the abstract state to determine if the target program satisfies given properties. With the development of model checking [2], mathematical logic has been widely applied to hardware and software model detection.

The Slither developed by Feist et al. [3], the Clairvoyance by Xue et al. [4], the ECFChecker by Grossman et al. [5], and the Sereum by Rodler et al. [6] utilize the rule-based detection method which checks whether a contract's control flow and data flow satisfy specific security properties based on predefined rules, such as preventing reentrancy attacks, airdrop attacks, information leaks, etc. These rules are typically formulated by security experts or developers. Nevertheless, the rule-based detection methods require manual rule definition, which may lead to false negatives and false positives. Additionally, this approach is often limited to detecting specific types of vulnerabilities rather than applicable to all threat scenarios.

Other approaches, such as the Oyente developed by Luu et al. [7], the Mythril by Mueller B [8], the Manticore by Mossberg et al. [9], the VerX by Permenev et al. [10], the RA by Chinen et al. [11], the EthBMC by Frank et al. [12], the Defectchecker by Chen et al. [13], and the BMC solving engine built into the Solidity compiler [14], use bounded model checking that enumerates the execution paths of a program and convert the paths into constraint expressions for verification. However, due to exponential growth in the number of program execution paths, the constraint-based model checking methods often cannot cover all possible paths when analyzing real-world contracts, thus leading to false negatives. VerX employs predicate abstraction to mitigate the state space explosion problem by abstracting the state space of the program into a set of disjoint regions, each composed of some states satisfying a set of predicates. This allows the state space to be partitioned into high-level predicates, reducing the number of states and improving the efficiency of model checking.

The VeriSol developed by Wang et al. [15], the ZEUS by Kalra et al. [16], the eTho by Schneidewind et al. [17], the SmartACE by Wesley et al. [18], the Vandal by Brent et al. [19], and the CHC solving engine built into the Solidity compiler [14] use the abstract refinement model checking method; first, the Solidity source code is coded in intermediate languages (i.e. Boogie [20], LLVM IR [21], or Datalog) through the front end of verification, and then the generated intermediate language is encoded into CHC constraint through the back end of verification before the CHC constraint is transformed into different SMT queries for verification.

The SmartPulse developed by Stephens et al. [22] also uses the abstract refinement model checking method. Based on the VeriSol tool, it replaces the verification backend of the intermediate language Boogie with the path abstraction technique developed by Heizmann et al. [23], i.e. Automizer. This enables the detection of liveness properties in an LTL constraint. Automizer abstracts the target model into a Floyd-Hoare automaton and utilizes path abstraction techniques to refine its states to significantly reduce the state space.

### 1.3 Main Contributions and Innovations of this Paper

This paper focuses on the threat detection of blockchain smart contracts. The following are the main contributions and innovations:

1. More complete recovery of control flow diagram based on recursive descent algorithm is achieved. Existing methods, such as Manticore, cannot completely recover the control flow diagram.
2. The cross-function reentrancy static detection method is implemented with a higher efficiency and accuracy based

on CHC constraint solving. Existing methods that perform cross-function reentrancy analysis are RA, eThor, Slither, Clairvoyance, ECFChecker, and Sereum. Specifically, RA uses constraint model checking which results in lower efficiency and eThor only detects whether the target contract can recursively initiate multiple cross-contract calls rather than checking if the contract state changes during the recursive calls. Slither and Clairvoyance employ rule-based detection methods and can only detect some threats. ECFChecker can only be used to judge single-path decisions at runtime, and Sereum can only perform detection based on taint analysis at runtime.

3. To address the shortcomings of existing methods, we propose a new smart contract threat detection method. It first uses recursive descent algorithm to recover the basic block and control flow diagram of contract code, and then performs static type inference on static single assignment (SSA) variables while combining the symbolic execution technique to encode the control flow diagram into CHC constraints. Finally, an automatic theorem prover based on abstraction refinement technology is used to verify the generated constraints and provide verification results for the target contract.
4. Corresponding vulnerability paths are automatically generated based on the detection results for automated vulnerability exploitation.

## 2 A Novel Smart Contract Threat Detection Method

In this paper, we propose a static analysis-based smart contract threat detection method CEIHorn, which can discover security vulnerabilities in smart contracts through static analysis. This method employs various techniques to optimize the representation of control flow diagrams for smart contract analysis. We utilize a recursive descent-based control flow diagram recovery algorithm to transform smart contract bytecode into control flow diagram. Additionally, a data flow-based static type inference algorithm is used to infer variable types in smart contracts for variable modeling in subsequent analyses.

We also propose some topological optimization strategies for the control flow diagram, including function-layered flow diagram and function flow diagram inlining. These optimization strategies can effectively reduce the size and complexity of the control flow diagram, thereby improving the efficiency of the constraint solver.

In order to improve the efficiency and accuracy of the security detection method, we propose a method to encode basic blocks and control flow diagrams as isomorphic CHC constraints. Using this method, the security properties and state transition constraints of smart contracts can be represented as CHC constraints, and then verified using a theorem prover based on abstract refinement. This enables fast static detection for common security threats.

In summary, the following subsections provide an extensive explanation of the smart contract threat detection method, CEIHorn. The method expands upon the techniques and optimization strategies used previously, which can be used effectively to discover the security vulnerabilities in smart contracts.

### 2.1 Recovery of the Smart Contract Control Flow Diagram

In this paper, we propose a control flow diagram recovery method for smart contracts based on the recursive descent algorithm which partitions smart contract bytecode into basic blocks and then uses a constraint solver to determine possible successors for each basic block, thereby establishing the control flow relationships between basic blocks.

Since there may be multiple successors at the end of a function's basic block, it is not possible to solely rely on internal constraints within the basic blocks to find the successor nodes. To address this issue, we employ the recursive descent algorithm. When encountering a basic block with uncertain successor nodes, the algorithm backtracks through the path stack to find the entry point of the current node's function and the amount of variation in the stack pointer. Based on this variation, it can identify all the return positions of the function containing the basic block, and effectively recover the smart contract control flow. As a result, a complete control flow diagram is obtained, providing support for subsequent program analysis.

In order to simplify the control flow diagram used for final model verification and reduce the topological complexity of the control flow diagram to improve the efficiency of the constraint solver, we propose three techniques: static type inference, function-layered flow diagram, and function flow diagram inlining. We use these techniques to transform and simplify the control flow diagram.

#### 2.1.1 Static Type Inference

In smart contracts, it is common to perform operations on the data stored in $storage$. However, $storage$ is stored in bytes, which means that for each variable to be read or written, a significant number of bitwise operations are required to locate its position in $storage$. This process incurs substantial overhead and reduces the efficiency of smart contract analysis.

To address this issue, we employ static type inference to determine variable types in the program. While analyzing the control flow, we analyze the data flow in the program to identify the type of variables. This technique categorizes all variables into a 256-bit vector abstract domain $\widehat{D_{B^{256}}}$ and a natural number abstract domain $\widehat{D_N}$, which effectively avoids unnecessary bitwise operations and improves the efficiency of program analysis.

#### 2.1.2 Function-Layered Flow Diagram

Smart contract code typically consists of multiple functions that may share the same code blocks. During the process of compilation optimization, these shared code blocks may be extracted and transformed into separate functions or subprograms for calls between different functions to improve code execution efficiency. This optimization technique is known as function extraction.

However, this technique may lead to a more complex topological structure of the generated control flow diagram, affecting the efficiency of the model checker. In order to address this issue, we utilize the function-layered flow diagram technique to separate shared code blocks into different function layers, which effectively reduces the complexity of the control flow diagram.

In the function-layered flow diagram technique, the first step is to abstract and partition the original control flow diagram of the smart contract code. The shared code blocks can be separated into different function layers. Then, in the solver, each function layer is processed separately to reduce the number of states that the solver needs to handle, as well as the branching conditions. This method can improve the efficiency of the model checker, enabling it to quickly identify potential vulnerabilities in the contract.

Fig. 3-4 shows an example of function-layered flow diagram, where the target contract has a total of 6 functions. Therefore, the function flow diagram is divided into 6 independent layers. The green nodes represent function entries and the purple nodes denote function exits.

*2.1.3 Function Flow Diagram Inlining*

We further reduce the topological complexity of the control flow diagram through function inlining which replaces function calls with the function body to reduce the costs caused by function calls. In smart contract optimization, function inlining is often used to reduce the costs of function calls.

By applying function inlining, we inline simple functions with numerous repetitive calls to their call sites, reducing the complexity of the topological structure of the control flow diagram. This, in turn, enhances the efficiency of the model checker, allowing the constraint solver to quickly identify potential vulnerabilities and errors in the smart contract.

*2.2 Smart Contract State Model*

$$status \coloneqq Execute \mid Stop \mid Revert \mid Call \mid Error \quad (3\text{-}1)$$
$$msg \coloneqq (caller, value, size, data) \quad (3\text{-}2)$$
$$state \coloneqq (pc, gas, status, stack, memory, msg) :: state \mid \epsilon \quad (3\text{-}3)$$
$$gstate \coloneqq (safe, reentry, path, storage, state) \quad (3\text{-}4)$$

To express the state of a smart contract with symbolic representation, we formally define the smart contract state model.

The above formulas represent the formal definition of the proposed smart contract state abstraction model. Specifically, $status$ represents the execution state and takes on five possible values, $Execute$, $Stop$, $Revert$, $Call$, and $Error$; $msg$ represents call information, including the caller, call value, call data size, and call data content; $state$ represents the virtual machine state and is defined as a recursive data structure, composed of a program counter, the remaining fuel, execution status, stack, $memory$, and the virtual machine state at the previous layer; $gstate$ represents the transaction chain state, including security status, reentrancy count, execution path, $storage$, and virtual machine state; $\widehat{D_N}$ and $\widehat{D_{B^{256}}}$ represent the abstract domains containing sets of natural numbers and 256-bit vectors, where the element "⊤" is added to estimate any legal elements in the set.

The smart contract standard execution model has two return states: $Stop$ and $Return$. For model checking, we merge $Stop$ and $Return$ and introduce four auxiliary states: $Execute$, $Revert$, $Call$, and $Error$. We also assume that when he $Revert$ state is reached, the entire transaction chain terminates rather than returning to the upper-level virtual machine state, and that the attackers initiate reentrant calls only once in a callback.

*2.3 Semantics of Smart Contract Bytecode*

$$Ins_{ADD} \coloneqq \frac{state(pc, stack) \land stack = a::b::tail}{state(pc+size_{ADD}, a+b::tail)} \quad (3\text{-}8)$$

$$Ins_{AND} \coloneqq \frac{state(pc, stack) \land stack = a::b::tail}{state(pc+size_{AND}, \top::tail)} \quad (3\text{-}9)$$

In order to express the state transitions of smart contracts using symbolic reasoning rules, we provide a formal definition for the semantics of smart contract bytecode.

To improve the efficiency of model checking, this paper performs constraint analysis only on some of the bytecode, while abstractly expresses other bytecodes. The above formulas show the constraints on the ADD and the AND instructions.

For the ADD instruction, we construct the above reasoning rules for constraint analysis on the program's execution behavior. Before execution, the program counter is at $pc$, and there are two elements $a$ and $b$ on the stack top. After execution, the program counter becomes $pc + \text{size}_{AND}$, and the two elements on the stack top are popped and replaced with a new element, $a + b$.

For the AND instruction, since the natural number logic does not support AND operations, we abstractly represent the execution behavior of the program by building the above reasoning rules. Before execution, the program counter is at $pc$, and there are two elements on the stack top. After execution, the program counter becomes $pc + \text{size}_{AND}$, and the two elements on the stack top are popped and replaced with a new element, ⊤.

*2.4 Smart Contract Execution State Transition*

$$\frac{gstate(safe, state) \land state(status) \land status = Stop}{gstate(safe, state(Error))} \quad (3\text{-}10)$$

$$\frac{gstate(reentry, state) \land state(pc, status) \land status = Stop}{gstate(reentry+1, state(0, Execute))} \quad (3\text{-}11)$$

$$\frac{gstate(state) \land state(status) \land status = Stop}{gstate(tail)} \quad (3\text{-}12)$$

$$\frac{gstate(reentry, state) \land state(status, stack) \land status = Call}{gstate(reentry+1, (0, Execute)::state(Execute, \top::tail))} \quad (3\text{-}13)$$

$$\frac{state(status, stack) \land status = Call}{state(Execute, \top::tail)} \quad (3\text{-}14)$$

In order to express the execution state transitions of smart contracts using symbolic reasoning rules, we provide a formal definition of the execution state transitions of smart contract bytecode.

The first three rules are used to check the $Stop$ execution state. Equation (3-10) identifies the safety status $safe$ when the execution state is $Stop$ and the transaction chain ends; if it is false, the execution state is modified to $Error$. Equation (3-11) indicates that when the execution state is $Stop$ and the transaction chain ends, the attacker initiates a new transaction chain. Equation (3-12) means that when the execution state is $Stop$ and the transaction chain is not yet finished, it returns to the virtual machine state in the previous layer.

The last two rules are used to check the $Call$ execution state. Equation (3-13) specifies that when the execution state is $Call$, the call target is the attacker's address, and the remaining $gas$ is greater than 2300, the attacker's contract code initiates another call to reenter the target contract, with the stack top value being ⊤. Equation (3-14) indicates that when the execution state is $Call$, and we assume that this call does not have a direct or indirect impact on the contract's safety status $safe$, it directly skips this call, and the stack top value is ⊤.

*2.5 Security Properties of Smart Contract*

Before model checking, we must first define the security property $\Phi$ of the model, namely providing the criteria for verifying if a model satisfies the requirements. After that, if we can find a model $\mathcal{M}$ that provides interpretations for all unexplained predicates in $\Phi$, i.e. $\mathcal{M} \models \Phi$, then we can prove that the model's security property $\Phi$ is established.

$$Inv_{Error} := \frac{gstate(safe, state) \wedge state(status) \wedge status = Error}{False} \quad (3\text{-}15)$$

Equation (3-15) checks if the $Error$ state is reachable. If it is, that means that when the transaction chain ends, the safety status $safe$ is $False$. This indicates that there is a transaction chain that can lead to unsafe operation, and this transaction chain is fully executed by the Ethereum virtual machine and recorded in the block information. In this case, the above reasoning rules cannot be satisfied, indicating that the safety property $\Phi$ of the model is not established.

After providing the reasoning rules for the $Error$ state, it is also necessary to present the reasoning rules related to the safety status $safe$ to complete the reasoning process. Based on the smart contract threat types discussed in Section 2, we propose reasoning rules for the safety status $safe$, i.e. $Inv_{Safe}$.

Based on the definitions of same-function and cross-function reentrancy attacks in Solidity, OpenZeppelin, and ConsenSys, we designed the reasoning rule $Inv_{Safe}$. If the target contract calls the attacker's contract, the attacker can modify the global state when reentering the target contract, and the target contract reads and writes the global state after the call, the target contract does not comply with the checks-effects-interactions paradigm and does not protect the critical path modifying the global state using a reentrancy guard. In this case, the recursive call path initiated by the attacker to the target contract cannot be unwound through tail recursion, indicating that the target contract may have a reentrancy vulnerability.

### 2.6 Constraint Encoding for Control Flow Diagram

The basic block is used as the fundamental unit to build CHC constraints for the basic unit. We first traverse all instructions within each basic block in turn to construct state constraints for each instruction according to the semantic rules of smart contract bytecode defined earlier and record the relative positions where each instruction reads and writes on the stack.

After processing all instructions within the basic block, we combine the state constraints before and after execution of each statement to obtain the internal state constraints of the basic block. We then construct state constraints between basic blocks based on their recorded read/write positions on the stack.

With the internal constraints $Inv_V$ for each basic block and the inter-block constraints $Inv_E$, we map the constraints inside the basic block to the nodes on the control flow diagram, and the constraints between the basic blocks to the directed edges on the control flow diagram to obtain the following CHC constraints that are isomorphic to the control flow diagram:

$$Inv_G = Inv_V \wedge Inv_E \quad (3\text{-}16)$$

### 2.7 Security Property Checking

After completing the constraint encoding of the control flow diagram to obtain the isomorphic control flow diagram constraint, i.e. $Inv_G$, we combine it with the previously defined state transition constraint $Inv_{Status} = Status_{Stop} \wedge Status_{Call}$, security property constraint $Inv_{Error}$, and security status transition constraint $Inv_{Safe}$, resulting in the following model security property:

$$\Phi = Inv_G \wedge Inv_{Status} \wedge Inv_{Error} \wedge Inv_{Safe} \quad (3\text{-}17)$$

We use the abstract refinement-based CHC constraint solving engine, Spacer, in the Z3 theorem prover to check the security property $\Phi$, and determine if the target smart contract has security issues based on the checking result of the solver.

If the solver determines that the security property $\Phi$ has a legal solution $\mathcal{M}$, it proves that the target contract does not violate the security property, thus demonstrating that the target contract is secure.

If the solver determines that there is no legal solution $\mathcal{M}$ for the security property $\Phi$ and provides a counterexample proof $\mathcal{P}$, it proves that the target contract violates the security property and may be insecure.

**Table 1.** Experimental Environment

| Environment | Details |
| --- | --- |
| Operating System | Kali Linux 2023.1 |
| Memory | 32 G |
| CPU | AMD Ryzen™ 5 5500 |
| Development Language | Python 3.11.2 |
| Theorem Prover | Z3 4.12.1 |

**Table 2.** Experimental Results

| Detection Results | CEIHorn | Mythril |
| --- | --- | --- |
| True Positive | 22 | 0 |
| False Negative | 19 | 45 |
| Detection Timeout | 4 | 0 |

### 2.8 Conclusions

We thoroughly explored the recovery and optimization of control flow diagram, smart contract state model, semantics of smart contract bytecode, smart contract execution state transition, and smart contract security properties and presented the encoding scheme of control flow diagram based on CHC constraints and obtained the complete model security properties. Additionally, a method for checking the security properties was proposed and examples of a legal solution and counterexample proofs were presented.

## 3 Experimental Results and Analysis

### 3.1 Experimental Setup

Due to the significant time and resource required in static analysis and the difficulty of annotating datasets for cross-function vulnerabilities, we focused solely on analyzing the contracts with vulnerabilities in real world. By testing contracts in real-world scenarios, we can better reflect the potential security issues in smart contracts and improve the credibility of detection results. In terms of detection results, we compare them with existing detection methods to demonstrate that the proposed CEIHorn method has a higher detection accuracy.

### 3.2 Experimental Environment

The experiments were conducted on the Kali Linux 2023.1 operating system using Python 3 as the development language. The control flow diagrams were visualized using Graphviz, and Z3 was used to check the generated models. For the complete experimental environment, see Table 1.

### 3.3 Experimental Results

We analyzed a total of 45 reentrancy attacks on smart contracts that occurred between 2016 and 2023, including the security of these contracts. The types of vulnerabilities included same-function reentrancy, cross-function reentrancy, cross-chain reentrancy, and read-only reentrancy. The proposed detection method, CEIHorn,was compared with the existing detection method Mythril, as shown in Table 2, with the detection timeout set to one hour. Mythril uses the command-line parameter "-m StateChangeAfterCall" to detect reentrancy vulnerabilities only.

According to the experimental results, the existing tool Mythril could not detect reentrancy vulnerabilities for real-world smart contracts. This was mainly due to Mythril's inability to fully recover control flow diagram and its lack of cross-path optimization techniques.

The proposed detection method, CEIHorn, exhibits better detection capabilities than Mythril. While Mythril can detect write operations at global state after external calls, it fails to handle recursive transaction chains generated by external transactions and cannot analyze the acquisition and release of reentrant locks for multiple functions, leading to both false negatives and false positives. In contrast, CEIHorn employs static type inference, the recursive descent-based control flow diagram recovery algorithm, and topological structure optimization strategies of control flow diagram to detect read and write operations at global state after external calls. By constructing recursive CHC constraints, the analysis of recursive transaction chains is implemented. This enables CEIHorn to simulate the state changes of reentrant locks and other global states in multiple recursive transactions, thereby enhancing the detection capabilities for various reentrancy vulnerabilities.

### 3.5 Conclusions

We conducted experiments to evaluate the performance and effectiveness of the proposed detection method, CEIHorn, in detecting reentrancy vulnerabilities.

First, we explained the experimental setup, in which CEIHorn is used to analyze the contracts with real-world vulnerabilities, and the existing detection method, Mythril, is used as comparison for performance and effectiveness evaluation.

Next, we introduced the experimental environment, including the operating system, CPU, and memory (including the 32 GB of memory) used in the experiment to ensure its stability and accuracy.

Finally, we compared CEIHorn with the existing detection method Mythril, and concluded that CEIHorn outperformed the latter in detecting reentrancy vulnerabilities. The experimental results demonstrate that CEIHorn has a high detection accuracy and a low false-positive rate. Compared to Mythril, CEIHorn exhibited better performance and effectiveness in detecting reentrancy vulnerabilities.

In conclusion, we validated the superiority and effectiveness of CEIHorn in detecting reentrancy vulnerabilities. This is important for ensuring the security and stability of Ethereum smart contracts.

## 4 Conclusions and Future Work

### 4.1 Scheme Conclusions

The main findings and accomplishments of this paper can be summarized as follows:

1. We used the recursive descent algorithm to recover the basic blocks and control flow diagram of contract code. Compared to existing methods that rely on path enumeration for recovery, our approach significantly improved the efficiency and completeness of control flow diagram recovery.
2. Static type inference was applied to SSA variables, reducing unnecessary bit operation constraints and lowering the inference overhead of constraint solver. Our detection method strikes a better balance between efficiency and precision than the existing methods that solely rely on bit vector for modeling of variables.
3. We encoded the control flow diagram as CHC constraints with symbolic execution techniques and utilized an automatic theorem prover based on abstract refinement to prove the generated constraints. Compared to the existing methods that generate path constraints through enumeration path, our approach improved the efficiency of threat detection by reusing constraint lemmas between different paths.
4. We propose a fast static detection method called CEIHorn for common security threats. Compared to the existing method Mythril, CEIHorn demonstrates higher detection accuracy. CEIHorn can detect both read and write operations in the global state after external calls using static type inference, the recursive descent-based control flow diagram recovery algorithm, and topological structure optimization strategies of control flow diagram. Furthermore, it can analyze recursive transaction chains by constructing recursive CHC constraints, while simulating the state changes of reentrant locks and other global states in multiple recursive transactions. This enhances the capabilities of detecting various reentrancy vulnerabilities.

*4.2 Future Work*

The work presented here has some limitations and we suggest potential areas for future research:

1. Although the model checking method based on abstract refinement significantly improved the efficiency of model checking and has been widely used in software vulnerability detection in the industry, invariant inference remains a challenging problem. Currently, Si et al. [24] proposed a deep learning (DL)-based invariant generation approach called Code2Inv. However, these approaches have some limitations and have not yet been put into practice. Therefore, future work can focus on determining how to use DL methods to generate invariants more quickly and accurately, thereby enhancing the efficiency of CHC constraint solving engine.

2. Regarding control flow diagram, we used a complete control flow diagram in the generated constraint model, which may lead to reduced efficiency in CHC constraint solving. In the future, we can combine the abstraction refinement process of function-layered flow diagram with that of the constraint model solving to reduce inference overhead for non-sensitive functions and thus improve the efficiency of detecting vulnerabilities in smart contracts.

3. In terms of state model design, we only modeled the execution state of a single contract without considering other contracts that may be involved in the execution state. This may result in a lower detection accuracy. Therefore, in the future, we can include other contracts that the target contract may call in the model and design corresponding inference rules for behaviors of mutual calls between multiple contracts, to more comprehensively detect threats for the target contract.

**References**

[1] X. Li, P. Jiang, T. Chen, X. Luo, Q. Wen, A survey on the security of blockchain systems, Future Gener. Comput. Syst. 107 (2020) 841-853. https://doi.org/10.1016/j.future.2017.08.020.

[2] E.M. Clarke, T.A. Henzinger, H. Veith, R. Bloem, Handbook of model checking, Springer, Cham, 2018.

[3] J. Feist, G. Grieco, A. Groce, Slither: a static analysis framework for smart contracts, in: 2019 IEEE/ACM 2nd International Workshop on Emerging Trends in Software Engineering for Blockchain (WETSEB), IEEE, Montreal, QC, 2019, pp. 8-15.

[4] Y. Xue, M. Ma, Y. Lin, Y. Sui, J. Ye, T. Peng, Cross-contract static analysis for detecting practical reentrancy vulnerabilities in smart contracts, in: Proceedings of the 35th IEEE/ACM International Conference on Automated Software Engineering, Association for Computing Machinery, Virtual Event, Australia, 2021, pp. 1029–1040.

[5] S. Grossman, I. Abraham, G. Golan-Gueta, Y. Michalevsky, N. Rinetzky, M. Sagiv, Y. Zohar, Online detection of effectively callback free objects with applications to smart contracts, Proc. ACM Program. Lang. 2 (2017) 1-28. https://doi.org/10.1145/3158136.

[6] M. Rodler, W. Li, G.O. Karame, L. Davi, Sereum: protecting existing smart contracts against re-entrancy attacks, arXiv preprint arXiv:1812.05934 (2018)

[7] L. Luu, D.H. Chu, H. Olickel, P. Saxena, A. Hobor, Making smart contracts smarter, in: Proceedings of the 2016 ACM SIGSAC Conference on Computer and Communications Security, Association for Computing Machinery, Vienna, Austria, 2016, pp. 254–269.

[8] B. Mueller, Smashing ethereum smart contracts for fun and real profit, HITBSecConf, Amsterdam, 2018.

[9] M. Mossberg, F. Manzano, E. Hennenfent, A. Groce, G. Grieco, J. Feist, T. Brunson, A. Dinaburg, Manticore: a user-friendly symbolic execution framework for binaries and smart contracts, in: 2019 34th IEEE/ACM International Conference on Automated Software Engineering (ASE), IEEE, San Diego, CA, 2019, pp. 1186-1189.

[10] A. Permenev, D. Dimitrov, P. Tsankov, D. Drachsler-Cohen, M. Vechev, VerX: safety verification of smart contracts, in: 2020 IEEE Symposium on Security and Privacy (SP), IEEE, San Francisco, CA, 2020, pp. 1661-1677.

[11] Y. Chinen, N. Yanai, J.P. Cruz, S. Okamura, RA: hunting for re-entrancy attacks in ethereum smart contracts via static analysis, in: 2020 IEEE International Conference on Blockchain (Blockchain), IEEE, Rhodes, Greece, 2020, pp. 327-336.

[12] J. Frank, C. Aschermann, T. Holz, ETHBMC: a bounded model checker for smart contracts, in: Proceedings of the 29th USENIX Conference on Security Symposium, USENIX Association, Berkeley, CA, 2020, pp. 2757-2774.

[13] J. Chen, X. Xia, D. Lo, J. Grundy, X. Luo, T. Chen, DefectChecker: automated smart contract defect detection by analyzing EVM bytecode, IEEE Trans. Softw. Eng. 48 (2022) 2189-2207. https://doi.org/10.1109/tse.2021.3054928.

[14] L. Alt, M. Blicha, A.E.J. Hyvärinen, N. Sharygina, SolCMC: solidity compiler's model checker, in: S. Shoham, Y. Vizel (Eds.), Computer Aided Verification, Springer International Publishing, Cham, 2022, pp. 325-338.

[15] Y. Wang, S.K. Lahiri, S. Chen, R. Pan, I. Dillig, C. Born, I. Naseer, K. Ferles, Formal verification of workflow policies for smart contracts in azure blockchain, in: S. Chakraborty, J.A. Navas (Eds.), Verified Software. Theories, Tools, and Experiments, Springer International Publishing, Cham, 2020, pp. 87-106.

[16] S. Kalra, S. Goel, M. Dhawan, S. Sharma, ZEUS: Analyzing Safety of Smart Contracts, Network and Distributed Systems Security (NDSS) Symposium, San Diego, CA, 2018.

[17] C. Schneidewind, I. Grishchenko, M. Scherer, M. Maffei, eThor: practical and provably sound static analysis of ethereum smart contracts, in: Proceedings of the 2020 ACM SIGSAC Conference on Computer and Communications Security, Association for Computing Machinery, Virtual Event, USA, 2020, pp. 621–640.

[18] S. Wesley, M. Christakis, J.A. Navas, R. Trefler, V. Wüstholz, A. Gurfinkel, Verifying solidity smart contracts via communication abstraction in SmartACE, in: Verification, Model Checking, and Abstract Interpretation, Springer International Publishing, Cham, 2022, pp. 425-449.

[19] L. Brent, A. Jurisevic, M. Kong, E. Liu, F. Gauthier, V. Gramoli, R. Holz, B. Scholz, Vandal: a scalable security analysis framework for smart contracts, arXiv preprint arXiv:1809.03981 (2018)

[20] M. Barnett, B.-Y.E. Chang, R. DeLine, B. Jacobs, K.R.M. Leino, Boogie: a modular reusable verifier for object-oriented programs, in: F.S. de Boer, M.M. Bonsangue, S. Graf, W.-P. de Roever (Eds.), Formal Methods for Components and Objects, Springer Berlin Heidelberg, Berlin, Heidelberg, 2006, pp. 364-387.

[21] C. Lattner, V. Adve, LLVM: a compilation framework for lifelong program analysis & transformation, in: International Symposium on Code Generation and Optimization, 2004. CGO 2004., IEEE, San Jose, CA, 2004, pp. 75-86.

[22] J. Stephens, K. Ferles, B. Mariano, S. Lahiri, I. Dillig, SmartPulse: automated checking of temporal properties in smart contracts, in: 2021 IEEE Symposium on Security and Privacy (SP), IEEE, San Francisco, CA, 2021, pp. 555-571.

[23] M. Heizmann, Y.-F. Chen, D. Dietsch, M. Greitschus, J. Hoenicke, Y. Li, A. Nutz, B. Musa, C. Schilling, T. Schindler, A. Podelski, Ultimate automizer and the search for perfect interpolants, in: D. Beyer, M. Huisman (Eds.), Tools and Algorithms for the Construction and Analysis of Systems, Springer International Publishing, Cham, 2018, pp. 447-451.

[24] X. Si, A. Naik, H. Dai, M. Naik, L. Song, Code2Inv: a deep learning framework for program verification, in: S.K. Lahiri, C. Wang (Eds.), Computer Aided Verification, Springer International Publishing, Cham, 2020, pp. 151-164.

**Figure Captions**
**Fig. 3-1.** Flow Diagram for Smart Contract Threat Detection Method, CEIHorn
**Fig. 3-2.** Example of Recovered Control Flow Diagram
**Fig. 3-4.** Example of Function-Layered Flow Diagram
**Fig. 3-5.** Example of Optimized Control Flow Diagram

**Figures**

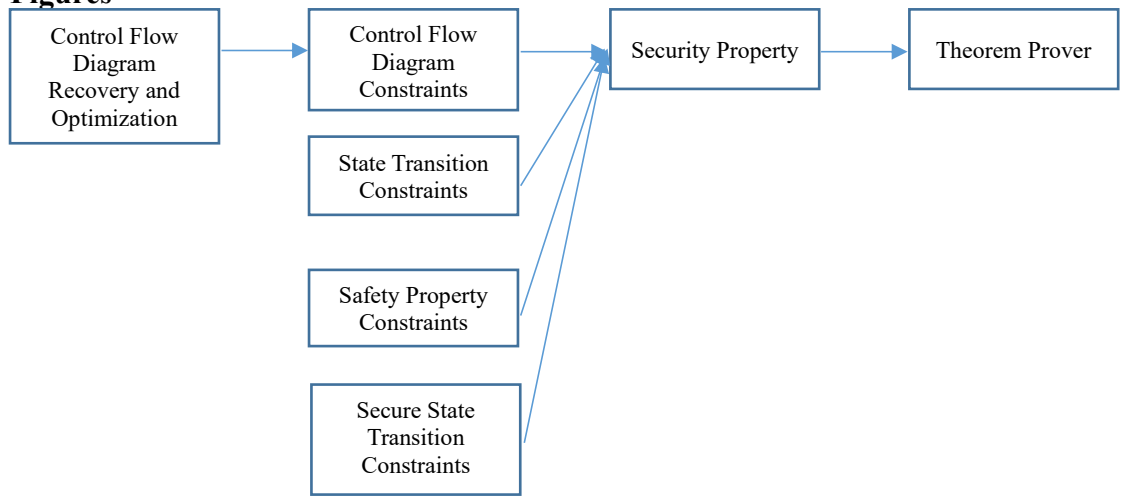

**Fig. 3-1**

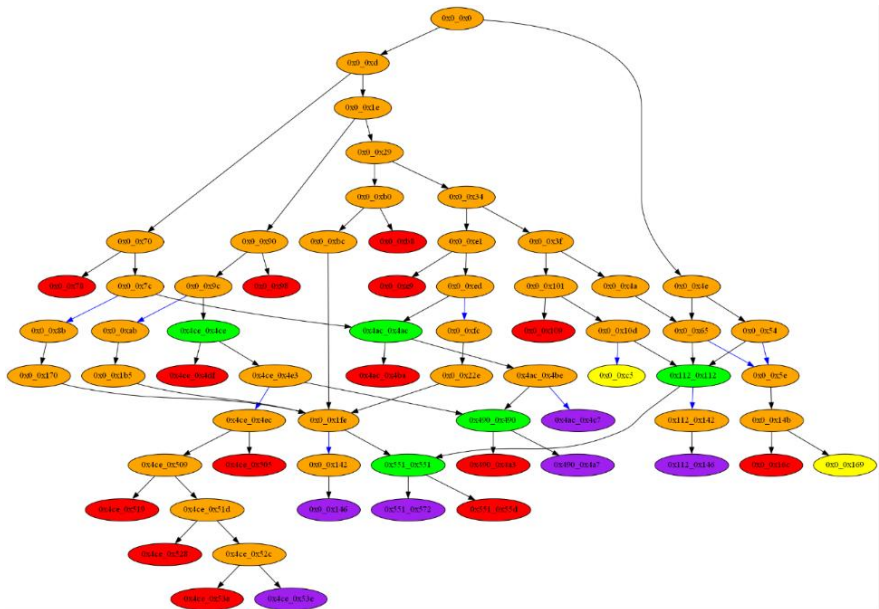

**Fig. 3-2**

**Fig. 3-4**

**Fig. 3-5**